\documentclass{iaus}
\usepackage{graphicx}

\title[Angular Momenta of  Preplanetesimals ] 
{Angular Momenta of Collided Rarefied Preplanetesimals}

\author[Sergei I. Ipatov ]   
{Sergei I. Ipatov $^{1}$
 \thanks{Present address: PO box 201536, Alsubai Est. for Scientific Studies, Doha, Qatar}
}

\affiliation{$^1$Space Research Institute, Moscow, Russia
\\ email: {\tt siipatov@hotmail.com}  }

\pubyear{2013}
\volume{293}  
\pagerange{119--126}
\jname{Formation, detection, and characterization of extrasolar habitable planets}
\editors{N. Haghighipour, ed.}
\begin{document}

\maketitle

\begin{abstract}
The angular momenta of rarefied preplanetesimals needed for formation of small-body binaries
can be obtained at collisions of preplanetesimals.
Trans-Neptunian objects, including trans-Neptunian binaries, could be formed from contracting rarefied preplanetesimals.

\end{abstract}

\firstsection 

\section{Introduction}

In recent years, the formation of rarefied preplanetesimals  was studied by several scientists
(e.g., \cite [Cuzzi et al. 2008]{Cuzzi_etal08}; 
\cite[Johansen et al. 2007, 2011, 2012]{Johansen_etal07};
\cite[Lyra et al. 2009)]{Lyra_etal09}.
\cite[Ipatov (2009, 2010a-b)] {Ipatov09} and
\cite[ Nesvorny et al. (2010)]{ Nesvorny_etal10}
 supposed that trans-Neptunian binaries were formed from rarefied preplanetesimals (RPPs). 
\cite[ Nesvorny et al. (2010)]{ Nesvorny_etal10} calculated contraction of RPPs and formation of binaries
supposing that RPPs got their angular momenta
when they formed from the protoplanet cloud.
\cite[Ipatov (2010a)] {Ipatov10} supposed that
 a considerable fraction of discovered trans-Neptunian binaries could acquire most of their angular momenta at collisions of RPPs.
He showed that the angular momenta acquired at collisions of RPPs moved before collisions in circular 
heliocentric orbits could have the same values as the angular momenta of discovered trans-Neptunian and asteroid binaries. 

In the present paper,  the angular velocities used by \cite[ Nesvorny et al. (2010)]{ Nesvorny_etal10} 
as initial data are compared with the angular velocities acquired for our model at a collision of two RPPs moving in circular orbits.
The frequency of collisions and mergers of collided preplanetesimals, and the formation of trans-Neptunian binaries are also discussed.

\section{Angular momentum of two colliding preplanetesimals}

 \cite[Ipatov (2010a)] {Ipatov10} obtained that 
the angular momentum of two  colliding RPPs
(with radii $r_1$ and $r_2$ and masses $m_1$ and $m_2$)  moved before the collision in circular heliocentric orbits  equals
$K_s=k_{\Theta}(G \cdot M_S)^{1/2}(r_1+r_2)^2m_1m_2(m_1+m_2)^{-1}a^{-3/2}$, 
where $G$ is the gravitational constant, 
 $M_S$ is the mass of the Sun, and 
 the difference in  semimajor axes $a$ of RPPs equals $\Theta(r_1+r_2)$. 
 At $r_a=(r_1+r_2)/a\ll \Theta$,
one can obtain $k_{\Theta} \approx 1-1.5\Theta^2$. 
$k_{\Theta}$ varies from -0.5 to 1.
The mean value of $|k_{\Theta}|$ equals 0.6. 
The values of $K_s$ are positive at $0<\Theta<0.8165$ 
and are negative at $0.8165<\Theta<1$. 

The angular velocity $\omega$ of the RPP
of radius $r=(r_1^3+r_2^3)^{1/3}$ and mass $m = m_1 + m_2$
 formed as a result of    a collision equals  $K_s$/$J_s$, where
 $J_s=0.4  \xi \cdot m \cdot r^2$ is the moment of inertia of a RPP,
  $\xi$=1 is for a uniform sphere considered by Nesvorny et al. (2010).   
I obtained that $\omega=2.5 k_{\Theta}  \xi^{-1} (r_1 + r_2)^2 r^{-2} m_1 m_2 (m_1 + m_2)^{-2 }\Omega$, 
where $\Omega = (G \cdot M_S  /a^3)^{1/2}$.
As $K_s/J_s \propto (r_1+r_2)^2/r^2$, then $\omega$ does not depend on $r_1$, $r_2$, and $r$ if
$(r_1+r_2)/r = $ $ $const. 
Therefore, $\omega$ will be the same at different values of $k_{r}$ if we consider RPPs with
radii $k_{r} r_{Hi}$, where $r_{Hi} \approx a (m_i/3 M_S)^{1/3}$ is the 
 radius of the  Hill sphere for mass $m_i$ ($m_1$, $m_2$, or $m$).
However, 
if at some moment of time after the collision of uniform spheres with radii $r_1$ and $r_2$, the radius
$r_c$ of a compressed sphere equals $k_{rc} r$, 
then (at $\xi$=1) the angular velocity of the compressed RPP is $\omega_c = \omega / k_{rc}^2$.
Below I consider $r_1$=$r_2$, $r^3$=$2 r_1^3$, $m_1$=$m_2$=$m/2$, and $\xi$=1. In this case, one has  
$\omega = 1.25 \cdot 2^{1/3} k_\Theta \Omega \approx 1.575 k_\Theta \Omega $.

\cite[ Nesvorny et al. (2010)]{ Nesvorny_etal10} made computer simulations
of  the contraction of preplanetesimals in the trans-Neptunian region.
They considered initial angular velocities of  preplanetesimals equal to
$\omega_{\circ}$=$k_{\omega} \Omega_{\circ}$, where 
$\Omega_{\circ} = (G m/r^3)^{1/2}$, 
$k_{\omega} =$ 
0.5, 0.75, 1, and 1.25. In most of their runs, $r$=$0.6 r_H$, where $r_H$ is the Hill radius of a RPP
of mass $m$. Also  $r$=$0.4 r_H$ and  $r$=$0.8 r_H$ were used.
Note that $\Omega_{\circ} / \Omega = 3^{1/2} (r_H / r)^{3/2} \approx 1.73 (r_H / r)^{3/2}$.

In the case of Hill spheres, considering $\omega$=$\omega_{\circ}$, we have
 $k_{\omega} = 1.25 \cdot 2^{1/3} 3^{-1/2} k_{\Theta} /\xi \approx 0.909  k_{\Theta} /\xi$. 
This relationship shows that
 it is possible to obtain the values of  
$\omega$=$\omega_\circ$ corresponding to $k_{\omega}$ up to 0.909
 at collisions of RPPs for $ k_{\Theta}$=$\xi$=1.
In the case of a collision of RPPs - Hill spheres and the subsequent contraction of the formed RPP to radius $r_c$,
the obtained angular velocity is $\omega_{rc} = \omega_H (r_H / r_c)^2$, 
where $\omega_H \approx 1.575 k_\Theta \Omega$.
For this RPP with radius $r_c$, we have
 $\omega_{\circ} =   (r_H / r_c)^{3/2} \Omega_{\circ H}$ (where $\Omega_{\circ H}= (G m/r_H^3)^{1/2}$)
and $\omega_{rc} / \omega_{\circ} \propto (r_H / r_c)^{1/2}$.
At $r_c / r_H = 0.6$, the collision of Hill spheres can produce $K_s$ corresponding to $k_{\omega}$ up to
$0.909 / 0.6^{1/2} \approx 1.17$.
\cite[ Nesvorny et al. (2010)]{ Nesvorny_etal10} obtained binaries or triples
only at $k_{\omega}$ equal to 0.5 or 0.75.
Therefore, one can conclude that the initial angular velocities of RPPs
that lead to formation of binaries
 can be obtained at collisions of RPPs.
Note that the values of $\omega$  at the moment of a collision are the same at collisions at different
values of $k_r$, but 
 $\omega_{rc}$$\propto$$r^2$ in the case of contraction of a RPP from $r$ to $r_c$. 

\section{ Frequency of collisions, mergers, and contraction of rarefied preplanetesimals}

The number of collisions of RPPs depends on the number of RPPs in the considered region,
on their initial sizes, and on the time dependences of radii of collapsing RPPs.
\cite[Cuzzi et al. (2008)]{Cuzzi_etal08} obtained the “sedimentation” timescale for RPPs 
to be roughly 30-300 orbit periods at 2.5 AU for 300 $\mu$m radius chondrules. 
Both smaller and greater times of contraction of RPPs were considered by other authors. 
According to
\cite[Lyra et al. (2009)]{Lyra_etal09},
the time of growth of Mars-size planetesimals from preplanetesimals consisted of boulders took place in five orbits.

It may be possible that a greater fraction of RPPs had not collided with other RPPs
before they contracted to solid bodies.
For an object with mass  $m_\circ = 6\cdot10^{17}$ kg $ $ $ \approx 10^{-7} M_E$ (where $M_E$ is the mass
of the Earth), e.g., for a solid object with diameter  $d$=100 km and density $\rho$$\approx$1.15 g cm$^{-3}$,
 its Hill radius equals $r_{H\circ} 
\approx 4.6\cdot10^{-5} a$.
For circular orbits separated by this Hill radius, the ratio of periods of motion of two RPPs
around the Sun is about $1+1.5 r_{Ha} \approx 1+7\cdot 10^{-5}$, where $r_{Ha}=r_{H\circ}/a$.
 In this case, 
the angle with a vertex in the Sun between the directions to the two RPPs changes by
 $2\pi \cdot 1.5 r_{Ha} n_r \approx 
0.044$  radians during $n_r$=100 revolutions of RPPs around the Sun.

Let us consider the planar disk consisted of $N$ identical RPPs with radii $r_\circ=r_{H\circ}$ and masses 
$m_\circ = 6\cdot10^{17}$ kg.
The ratio of the distances from the Sun to the  edges of the disk is supposed
to be equal to $a_{rat}$=1.67 (e.g., for a disk from 30 to 50 AU).
For $N$=$10^7$ and $M_\Sigma = m_\circ  N = M_E$,
a RPP can collide with another RPP when their semimajor axes differ by not more than
$2 r_{H\circ}$, i.e., the mean number of RPPs which can collide with the RPP is about
$2N \cdot r_{Ha}(a_{rat}+1) /(a_{rat} -1)  \approx 3.7 \cdot 10^3$ or
$4/3 \cdot N \cdot r_{Ha}(a_{rat}^2+a_{rat}+1) /(a_{rat}^2 -1)  \approx   1.9 \cdot 10^3$
if the number of RPPs is proportional to $a$ or $a^2$, respectively.
The mean number $N_c$ of collisions of the RPP during $n_r$ revolutions around the Sun can be estimated as
$ 1.5 r_{Ha} n_r N_m$. At $N_m \approx 3 \cdot 10^3$ and $r_{Ha} \approx 4.6 \cdot 10^{-5}$, we have $N_c \approx 0.2 n_r$.
 $N_c$ is proportional to $N \cdot r_{Ha}^2 \propto N \cdot m_\circ^{2/3} \propto M_{\Sigma} \cdot  m_\circ^{-1/3}$.
In calculations by Nesvorny et al. (2010), the binary systems  formed from RPPs in 100 yr (i.e. in 0.6 of the orbital period at 30 AU). 
At time equal to 0.6 of the orbital period, the fraction of collided RPPs for our above model is about 0.3,
 which is in accordance with the fraction of classical 
trans-Neptunian objects (TNOs)
 with satellites among discovered classical TNOs. 
Some collisions were tangent and did not result in a merger.
RPPs contracted with time. Therefore, the real number of mergers can be  smaller than that for the above estimates.

Densities of RPPs could be very low, but their relative velocities $v_{rel}$
 at collisions were also very small. The velocities  $v_{rel}$ were smaller than 
the escape velocities on the edge of the Hill sphere of the primary 
\cite[(Ipatov 2010a)] {Ipatov10}.
 If  collided RPPs are much smaller than their Hill spheres, and if their heliocentric
 orbits are almost circular before the encounter, then the velocity of the collision does not differ much
 from the parabolic velocity $v_{par}$ at the surface of the primary RPP (with radius $r_{pc}$).
 Indeed, $v_{par}$ is proportional to    $ r_{pc}^{-0.5}$. 
Therefore, collisions of RPPs could result in a merger
 (followed by possible formation of satellites) at any $r_{pc}<r_H$.
\cite[Johansen et al. (2007)]{Johansen_etal07} 
 determined that the mean free path of a boulder inside a cluster -- preplanetesimal is shorter than
 the size of the cluster. This result supports the picture of mergers of RPPs. In calculations made by
\cite[Johansen et al. (2011)]{Johansen_etal11}, 
 collided RPPs merged.  Given a primary of mass $m_p$ and a much smaller secondary,
 both in circular heliocentric orbits, one can obtain that the ratio $v_{\tau}/v_{esc-pr}$ of tangential
 velocity of an encounter up to the Hill radius to the escape velocity on the edge of the Hill sphere is
proportional to $m_p^{-1/3}\cdot a^{-1}$. 
Therefore, collided RPPs are more likely to merge when the primary is more massive and located
 farther from the Sun. 

If  a RPP  got its angular momentum at the collision of two RPPs at $a$=1 AU,  $k_\Theta$=0.6, $k_r$=1,
and $m_1$=$m_2$, then
the period  $T_{s1}$ of rotation of the planetesimal of density $\rho$=1 g cm$^{-3}$
formed from the RPP with intial radius $ r_H$ is
 $\approx$0.5 h, i.e., is smaller than the periods (3.3 and 2.3 h) at which
velocity of a particle on a surface of a rotating spherical object at the equator
is equal to the circular and escape velocities, respectively.
$T_{s1}$ is proportional to $a^{-1/2}\rho^{-2/3}$. Therefore,  $T_{s1}$
and the fraction of the mass of the RPP 
that could contract to a solid core are smaller for greater $a$.
For those collided RPPs that were smaller than their Hill spheres and/or differ in masses,
$K_s$ was smaller (and $T_{s1}$ was greater) than for the above case with  $k_r$=1 and $m_1$=$m_2$.

\section{Formation of trans-Neptunian objects and their satellites}

The above discussion can explain why a fraction of discovered binaries is greater for greater distances 
between the place of their origin and the Sun, 
and why the typical mass ratio of secondary to primary is greater for 
TNOs than for asteroids. 
Among 477 main-belt asteroids, Pravec et al. (2012) found 45 binaries (i.e. the fraction of the binaries is 0.094).
The binary fraction in the minor planet population is about 0.29 for cold classical TNOs and is 0.1 for all other TNOs 
\cite[(Noll et al. 2008)]{Noll_etal08}. 
 Note that TNOs moving in eccentric orbits (mentioned above as ``other TNOs") are thought to have been 
formed near the giant planets, closer to the Sun than classical TNOs 
(e.g., 
\cite[Ipatov 1987]{Ipatov87};
\cite[Levison \& Stern 2001]{LevisonStern01};
\cite[Gomes 2003]{Gomes03}).
Most of rarefied preasteroids could turn into solid asteroids before they collided with other preasteroids. 
Some present asteroids 
can be debris of larger solid bodies, and the formation of many binaries with primaries  with diameter $d<100$ km
can be explained by other models (not by contraction of RPPs). 
In the considered model of binary formation, two colliding RPPs originate at almost the same distance from the Sun.
 This point agrees with the correlation between the colors of primaries and secondaries obtained by 
\cite [Benecchi et al. (2009)] {Benecchi09}
 for trans-Neptunian binaries. In addition, 
the material within 
the RPPs could have been mixed before the binary components formed. 

\cite[Nesvorny et al. (2010)]{ Nesvorny_etal10} supposed that the angular velocities they used were produced during
 the formation of RPPs without collisions.
They noted that simulations by Johansen et al. seem to generally indicated prograde rotation.
Angular momenta of some observed trans-Neptunian binaries are negative.
The model of collisions of RPPs explains negative angular momenta of some observed binaries, as
about 20\% of collisions of RPPs moving in almost circular heliocentric orbits lead to  retrograde rotation. Note that if
 all RPPs got their angular momenta at their formation without mutual collisions,
then the angular momenta of minor bodies without satellites and those with satellites
 could be similar (but actually they differ considerably).
In my opinion, those RPPs that formed TNOs with satellites acquired most of their 
angular momenta at collisions.

The formation of classical TNOs from RPPs could have taken place for a small total mass 
of RPPs in the trans-Neptunian region, even given the present total mass of TNOs. 
Models of formation of TNOs from solid planetesimals (e.g., 
\cite[Stern 1995] {Stern95})
require a massive primordial belt and small ($\sim$0.001) eccentricities during the process of accumulation. 
However, the gravitational interactions between planetesimals during this stage could have increased 
the eccentricities to values far greater than those mentioned above (e.g., \cite[Ipatov 2007] {Ipatov07}).
This increase testifies in favor of the formation of TNOs from  RPPs.

\section{Conclusions}
The angular momenta of rarefied preplanetesimals needed for formation of small-body binaries
can be obtained at collisions of preplanetesimals.
Trans-Neptunian objects, including those with satellites, could be formed from contracting rarefied preplanetesimals.
The fraction of preplanetesimals collided with other preplanetesimals during their contraction
can be about the fraction of small bodies of diameter $d > 100$ km with satellites
(among all such small bodies), i.e., it can be about 0.3 in the trans-Neptunian belt.


\begin{thebibliography}{}

\bibitem[Benecchi et al. (2009)] {Benecchi09}
{Benecchi, S.D., Noll, K.S., Grundy, W.M., Buie, M.W., Stephens, D.C., \& Levison, H.F.} 2009,
\textit{Icarus}, 200, 292-303

\bibitem[Cuzzi et al.(2008)]{Cuzzi_etal08} {Cuzzi, J.N., Hogan, R.C., \& Shariff, K.} 2008, \textit{ApJ},  687, 1432-1447

\bibitem[Gomes (2003)]{Gomes03}
{Gomes, R.S.} 2003,
\textit{Icarus}, 161, 404-418

\bibitem[Ipatov (1987)]{Ipatov87}
{Ipatov, S.I.} 1987,
\textit{Earth, Moon, \& Planets}, 39, 101-128 

\bibitem[Ipatov (2007)] {Ipatov07}
{Ipatov, S.I.} 2007,
\textit{LPS} XXXVIII, Abstract \#1260

\bibitem[Ipatov (2009)] {Ipatov09}
{Ipatov, S.I.} 2009, \textit{LPS} XL, Abstract \#1021

\bibitem[Ipatov (2010a)] {Ipatov10a}
{Ipatov, S.I.} 2010a,\textit{MNRAS},  403, 405-414

\bibitem[Ipatov (2010b)] {Ipatov10b}
{Ipatov, S.I.} 2010b, in J.A. Fernandez, D. Lazzaro, D. Prialnik, R. Schulz (eds.),
 \textit{Icy bodies in the Solar System}, Proc. IAU Symp. No 263  (Cambridge University Press), p. \ 37-40

\bibitem[Johansen et al.(2007)]{Johansen_etal07}
{Johansen, A., Oishi, J.S., Mac Low, M.-M., Klahr, H., Henning, T.,  \& Youdin, A.} 2007,
\textit{Nature}, 448, 1022-1025

\bibitem[Johansen et al.(2011)]{Johansen_etal11}
{Johansen, A., Klahr, H., \& Henning, T.} 2011,
\textit{A\&A}, 529, A62

\bibitem[Johansen et al.(2012)]{Johansen_etal12}
{Johansen, A., Youdin, A.N.,  \& Lithwick, Y.} 2012,   
\textit{A\&A},
537, A125

\bibitem[Levison \& Stern (2001)]{LevisonStern01}
{Levison, H.F. \& Stern, S.A.} 2001, 
\textit{AJ}, 121, 1730-1735

\bibitem[Lyra et al.(2009)]{Lyra_etal09}
{Lyra, W., Johansen, A., Zsom, A., Klahr, H., \& Piskunov, N.} 2009, 
\textit{A\&A}, 497, 869-888

\bibitem[ Nesvorny et al. (2010)]{ Nesvorny_etal10}
{Nesvorny, D., Youdin, A.N., \& Richardson, D.C.} 2010, \textit{AJ}, 140, 785-793 

\bibitem[ Noll et al. (2010)]{ Noll_etal10}
{Noll, K.S., Grundy, W.M., Stephens, D.C., Levison, H.F., \& Kern, S.D.} 2008, 
\textit{Icarus}, 194, 758-768

\bibitem[ Pravec et al. (2012)]{ Pravec_etal12}
{Pravec, P. et al.} 2012, \textit{Icarus}, 218, 125-143

\bibitem[Stern (1995)] {Stern95}
{Stern, S.A.} 1995,
\textit{AJ}, 110, 856-868 

\end{thebibliography}
\end{document}